# Applications and challenges of Reconfigurable Intelligent Surface for 6G networks[1][2]

ZHAO Yajun[1,2], JIAN Mengnanx[1,2]

(1. ZTE Corp., Beijing 100192, China;

2. State Key Laboratory of Mobile Network and Mobile Multimedia Technology, Shenzhen 518055, China.)

**Abstract:** Reconfigurable intelligent surface has attracted the attention of academia and industry as soon as it appears because it can flexibly manipulate the electromagnetic characteristics of wireless channel. Especially in the past one or two years, RIS has been developing rapidly in academic research and industry promotion and is one of the key candidate technologies for 5G-Advanced and 6G networks. RIS can build a smart radio environment through its ability to regulate radio wave transmission in a flexible way. The introduction of RIS may create a new network paradigm, which brings new possibilities to the future network, but also leads to many new challenges in the technological and engineering applications. This paper first introduces the main aspects of RIS enabled wireless communication network from a new perspective, and then focuses on the key challenges faced by the introduction of RIS. This paper briefly summarizes the main engineering application challenges faced by RIS networks, and further analyzes and discusses several key technical challenges among of them in depth, such as channel degradation, network coexistence, network coexistence and network deployment, and proposes possible solutions.

**Keywords:** 6G, Reconfigurable intelligent surface, Channel rank reduction; Multi-network coexistence; Multi-user multiplex; Multi-cell coexistence; Network deployment

---

## 0 Introduction

With the large-scale deployment of 5g commercial network, the number of 5G users and the demand for mobile communication services are growing rapidly. At the 3GPP PCG#46 meeting held on April 27-28, 2021, "5G -Advanced" was officially adopted to identify the standard version after 5G rel-18, and the Rel-18 Workshop [1] was held from June 27 to July 2 to discuss the requirements and candidate technical characteristics of 5G rel-18 and subsequent versions, marking that 5G standardization is about to enter the second stage [2]. Compared with the first phase of 5G, the second phase of 5G standard (5g-advanced) puts forward higher requirements [3]. At the same time, 6G demand exploration and potential key research work are also in full swing, putting forward new business and scenario requirements and higher requirements for technology [4].

For a long time, it has been people's dream to regulate electromagnetic waves at will. The emergence of Maxwell's equations has led to the rapid growth of human's ability to control electromagnetic waves. However, due to the relatively fixed electromagnetic parameters of natural materials, people's control over electromagnetic waves is limited to transmitters and receivers. In recent years, reconfigurable intelligent surface (RIS) has attracted extensive attention in the academic and industry since it can flexibly manipulate the electromagnetic characteristics in the channel environment. Especially in the past one or two years, RIS has developed rapidly in academic research and industrial promotion, and is considered to be one of the key candidate technologies of 5G-Adv [5] [6] [7] [8] and 6Gnetwork [9] [10]. RIS is usually composed of a large number of carefully designed electromagnetic units. By applying control signals to the adjustable elements on the electromagnetic units, the electromagnetic properties of these electromagnetic units can be dynamically controlled, so as to realize the active and intelligent regulation of space electromagnetic waves in a programmable way, and form an electromagnetic field with controllable amplitude, phase, polarization and frequency. As a two-dimensional implementation of metamaterials, RIS naturally has the characteristics of low cost, low complexity and easy deployment. It has the opportunity to solve the new needs and challenges faced by wireless networks in the future. RIS introduces the wireless communication network, which makes the wireless propagation environment from passive adaptation to active control, so as to build a smart radio environment (SRE). The concept of SRE introduces a new communication theory viewpoint for wireless communication system and provides a new opportunity for optimization [11].

Existing RIS researches mainly focus on the new challenges faced by classical communication problems after the introduction of RIS, such as channel estimation, beamforming, etc., and focus on single user or simple multi-user scenarios. Relevant literature has carried out in-depth theoretical analysis on the above problems, and put forward corresponding solutions. However, the introduction of RIS may build a new network paradigm, which not only brings new possibilities to the future network, but also leads to many new technical and engineering application challenges. This paper will focus on the in-depth analysis and discussion of these new challenges, and put forward possible solutions or research ideas.

Firstly, this paper introduces the main aspects of RIS enabling future wireless communication network, including reconstructing wireless channel, enhancing communication capability, enabling wireless network to provide more network possibilities and realizing end-to-end full link intelligence. Then, it briefly summarizes the main engineering application technical challenges faced by RIS network, deeply analyzes and discusses several key technical challenges such as channel rank reduction, inter network coexistence, intra network coexistence and network deployment, and puts forward possible solutions or research ideas. Finally, the paper gives a summary and prospect.

## 1 RIS enabling future wireless communication network

As an artificial material of sub wavelength two-dimensional surface, RIS shows many excellent technical characteristics compared with traditional communication technology, which can well enable future wireless networks. Its typical technical characteristics mainly include four aspects: (1) passive, no thermal noise, low power consumption, meeting the requirements of green communication; (2) Low cost, no need for high-cost devices such as mixer, digital to analog converter and power amplifier; (3) Easy to deploy, scalable and lightweight design makes it easy to install and disassemble; (4) Electromagnetic waves can be reconstructed at any point on the continuous surface, which can form any shape surface to adapt to different application scenarios and support higher spatial resolution. In addition, the intelligent hypersurface has full band response characteristics, can support full duplex transmission, and generally has large area, high receiving energy, and is convenient for intensive deployment, which has the natural advantages of engineering practice [12]. Based on the above excellent technical characteristics of RIS, it has positive significance in solving the problem of non line of sight transmission, expanding coverage, reducing electromagnetic pollution, environmental perception and positioning, realizing green communication, and can be used as the key supporting technology of communication perception integration in the future.

1.1 Reconstruct the wireless channel and enhance the communication capability

There are two problems in the wireless channel of traditional networks. First, the channel capacity is limited by the characteristics of natural propagation channel, and can only approach the upper limit of channel capacity under the constraint of natural propagation channel by optimizing transceiver algorithm; Second, the large-scale and small-scale channel parameters of the transmission channel are random and time-varying, which is difficult to avoid the channel estimation deviation, resulting in channel matching deviation regardless of link adaptation or beamforming, and there is a gap between the actual achievable capacity and the ideal channel capacity.

The introduction of RIS will bring new changes in channel characteristics, have the opportunity to realize the reconstruction of wireless channel and break through the limitation of natural wireless propagation channel in traditional network. Its changes to channel characteristics can be mainly reflected in the following three aspects:

(1) Properly deploy a sufficient number of RIS and use its abnormal regulation ability to control the main propagation path of electromagnetic wave, greatly change the intensity distribution characteristics of electromagnetic wave, and realize the control of channel large-scale parameters. For example, millimeter wave beam is easy to block, RIS can be properly deployed for scattering regulation, large phase adjustment of signal propagation path and coverage reconstruction. Reasonable deployment and regulation of RIS can strictly restrict the coverage area, so as to reduce the interference coupling relationship between cells.

(2) The ubiquitous RIS performs adaptive dynamic regulation to realize the dynamic on-demand regulation of the

delay, amplitude, phase and polarization mode of the main scatterers/multipaths of the signal, so as to control the small-scale parameters of the channel, so as to suppress the multipath effect and change the spatial characteristics of MIMO channel. For example, RIS can track the movement of UE by regulating the regulation of reflected signal, so as to reduce the probability of channel deep fading as much as possible, including small-scale deep fading caused by multipath and large-scale deep fading caused by shadow.

(3) The multipath channel parameters are dynamically adjusted, and the small-scale parameters of the channel change from random dynamic time to controllable to a certain extent, so that the channel parameters can be estimated and predicted more accurately.

It can be seen from the above that due to the characteristics of low cost, low power consumption and simple and easy deployment, RIS has the opportunity to be widely deployed. The ubiquitous RIS has the opportunity to change the natural propagation channel and break the inherent capacity limit of the natural propagation channel by changing the channel, so as to improve the channel capacity. Its core capability comes from the intelligent control and change of natural wireless channel. Through the regulation of reflection phase, amplitude and polarization mode, the wireless channel is changed from a highly random time-varying natural channel to an artificial channel with a certain degree of certainty and controllability

## 1.2 Enable wireless networks to provide more possibilities

RIS introduces wireless network. Because of its new technical characteristics, it not only enhances the traditional communication ability, but also provides more possibilities for wireless network in the future. The possible applications of RIS are discussed in many literatures, including typical applications such as high-precision perception, location, simultaneous interpreting, backscatter, secure communication, and reduction of electromagnetic pollution.

Without losing generality, take RIS supporting perceptual positioning applications as an example. The traditional cellular network provides wireless positioning function. Its positioning accuracy is limited by the deployment location of base stations, the number of base stations and the size of antenna array, and its space and angle identification ability is limited. Compared with traditional transceivers, the advantages of using RIS for electromagnetic sensing include: (1) RIS is easy to deploy on a large scale and can realize sensing and communication without blind area （2) A large number of units of RIS can collect rich information during channel sensing, so as to obtain high-precision and fine-grained environment sensing results （3) The large amount of data information obtained can be applied to data-driven artificial intelligence technology, so as to mine more comprehensive and accurate environmental information. Therefore, RIS has the opportunity to become one of the key supporting technologies to support the integration of communication perception in the future.

## 1.3 Realize end-to-end full link intelligence

In recent years, AI is used to enhance the traditional wireless communication system. However, the existing research mainly discusses the intelligence of the transmitter and receiver, and the wireless channel still needs to passively adapt to the natural propagation environment. RIS has the opportunity to build an intelligent radio electromagnetic environment through its ability to flexibly and abnormally regulate radio wave transmission. That is, the introduction of RIS enables the wireless network to further realize the intelligent control of wireless channel on the basis of realizing the intelligence of wireless system receiving and transmitting ends, so as to have the opportunity to build an end-to-end intelligent wireless system covering transmitting end, wireless channel and receiving end, and support the realization of a truly "intelligent ubiquitous" and "intelligent endogenous" future 6G network[13]

In conclusion, RIS can build an intelligent and controllable wireless communication environment, break through the limitations of traditional communication networks, and provide more possibilities for future wireless networks on the basis of enhancing traditional communication capabilities.

## 2 Key Challenges and Solutions

The introduction of RIS may build a new network paradigm, which brings new possibilities to the future network, but also leads to many new technology and engineering application challenges. Table 1 summarizes the main engineering application technology challenges facing RIS network. At present, RIS engineering application technology research is still in the initial stage, involving more technical problems, limited to space. This paper will only analyze and discuss the key technical challenges in channel rank reduction, network coexistence, network coexistence, network deployment and other key technical challenges, and propose the possible solutions or research ideas. We will further analyze and discuss other technical challenges not launched in this paper at an appropriate time. Here, we will briefly list them for the reference of researchers.

Tab.1 Challenges faced by RIS networks in technology and engineering applications

| Key Challenges and Solutions | | Descriptive |
|---|---|---|
| Impacts on wireless channel characteristics | | The introduction of RIS makes the original natural uncontrollable electromagnetic propagation environment into an artificially controllable electromagnetic propagation environment. The active regulation of the electromagnetic propagation environment may bring a new channel characterization paradigm.<br>① It effectively enhances the signal, but also reduces the degree of spatial freedom and reduces the channel rank<br>② Large scale characteristic changes, such as the multiplicative relationship between NB-RIS-UE path loss and segmented channel distance<br>③ Significant near-field characteristics and significant spatial nonstationary characteristics<br>④ Mutual coupling problem caused by compact structure (basic unit spacing less than half wavelength) |
| Impact on different signal types | Optimal regulation of "target signal" | Target signal: optimize the propagation channel and control the propagation of target signal.<br>① Multiple business types coexist. For example, eMBB, URLLC, mMMTC and so on, different optimization objectives.<br>(2) UE of different scenarios, for example, the UE of cell center needs multiplexing gain, and UE of cell edge needs to enhance signal.<br>③ Multipath signal regulation. The incident angles of signals of different paths on the RIS surface are different, and there may be great differences. How to optimize and control the multipath signals with different incident angles is a challenge.<br>④ Ultra wideband, such as THz band, the dispersion problem of beam regulation. |
| | Abnormal regulation of "non target signal" | The "target signal" is enhanced by adjusting the amplitude, phase and polarization mode of electromagnetic wave, and the "non target signal" is also adjusted unexpectedly.<br>① Intra-network coexistence: multi-user multiplexing and multi cell coexistence<br>②Inter-networks Coexistence: spectrum sharing between networks (co channel coexistence), different frequencies between networks (adjacent channel coexistence) |
| Network deployment and optimization complexity | | The introduction of RIS brings a new network paradigm to the future network, but it also leads to the complexity of network deployment and optimization.<br>① Multiple transport nodes share the same RIS<br>② Cooperative scheduling when a transmission node uses multiple RIS at the same time<br>③ Control link between transmission node and RIS<br>④ The challenge of network topology planning and optimization. RIS enhances the extended signal coverage area, which may |
| | | break the traditional strictly divided sector coverage characteristics and bring complexity to network planning and optimization<br>⑤ Station site selection and power supply. The simplicity, low cost and low power consumption of RIS make it possible for RIS to be deployed more generally. However, it still faces the problem of site selection, and ubiquitous deployment will also bring challenges to power supply<br>⑥ Different scenes require different forms of RIS. For example, transmission type and reflection type; From the perspective of dynamic regulation, including static weighting, semi-static regulation, dynamic regulation and so on; Active unit with or without sensing capability; Whether there is power amplification capability, etc. It can be pin tube, liquid crystal, varactor and other materials with different cost and technical characteristics |
| Influence of Engineering deviation | ① | Adjust the phase quantization error. For example, 1 bit quantization, 2 bit quantization |
| | ② | Deviation calibration, including processing deviation and calibration, aging / environmental impact and other system drift measurement calibration, etc. For example, the RIS panel has been calibrated in the factory, but there is system drift in the actual network, which needs accurate measurement and compensation |
| | ③ | Codebook quantization error |
| | ④ | CSI feedback overhead delay |
| Other challenges that engineering applications may face | ① | Synchronization problems, including synchronization between base station and RIS, synchronization between RIS, synchronization between RIS and UE, etc |
| | ② | Multi band and cross band coexistence |
| | ③ | Station site requirements, power supply, network maintenance, etc. in deployment |
| | ④ | Weight problem. In addition to the RIS panel, the RIS overall structure also needs necessary support and protection structures, and it is difficult to reduce the weight of necessary auxiliary structures other than the main structure |
| Design of typical communication process in RIS network | ① | UE access process design. RIS measurement, discovery, random access, transmission and other processes, such as target signal identification and regulation; Measurement and discovery of RIS panel |
| | ② | Resource management and scheduling process. It mainly refers to the management and scheduling of RIS resources, such as the selection of RIS channel and NB-UE direct channel, and the scheduling / selection of RIS in multiple RIS scenarios |
| | ③ | The impact of RIS introduction on the handover process (mobility management). For example, switching between RIS scenarios, switching between different RIS, and switching between different frequency bands |

## 3 Influence of wireless channel characteristics - channel rank reduction

Compared with the traditional wireless communication system, the introduction of RIS will have a significant impact on the characteristics of wireless channel. For example, the large-scale array / antenna aperture makes the near-field characteristics of the channel significant; Super dense array (<0.5 $\lambda$) brings mutual coupling problem; The possible nonlinear impact on electromagnetic wave propagation and the possible changes of new modes caused by wide deployment on wireless propagation environment. In addition, the introduction of RIS makes the original natural uncontrollable electromagnetic propagation environment into an artificially controllable electromagnetic propagation environment. The active regulation of the electromagnetic propagation environment may bring a new channel characterization paradigm. Based on our limited understanding, it is found that the existing research mainly focuses on the large-scale characteristics of RIS [14], channel estimation [15], beamforming gain [16], etc. no literature has been found to conduct in-depth analysis and Research on the impact of the introduction of RIS on MIMO spatial multiplexing performance.

RIS enabled channels include two segmented subchannels, NB-RIS and RIS-UE. Typically, the channels between NB-RIS are Los channels or NLOS low rank channels, so even if the channels between RIS-UEs are high rank channels with rich scattering, the rank of NB-RIS-UE cascaded channels is still limited by the channels between NB-RIS. In this section, we will discuss in depth the reduction of spatial freedom of wireless channel after the introduction of RIS, which leads to the reduction of channel rank.

3.1 System Model

For a RIS assisted wireless communication system, consider an NB with $M$ antennas, an RIS with $N$ antenna units, and a UE with $U$ antennas. Let $H_{\text{nb-ue}} \in C^{U \times M}$ represent the direct channel between NB and UE, $G_{\text{nb-ris}} \in C^{N \times M}$ represent the channel between NB and RIS, and $H_{\text{ris-ue}} \in C^{U \times N}$ represent the channel between UE and RIS.

There are two scenarios for the overall propagation channel between NB and UE, which can be expressed as follows:

Scenario 1: both direct channels and channels passing through RIS are included

$$H_T = H_{\text{ris-ue}} \beta \Theta_{\text{ris}} G_{\text{nb-ris}} + H_{\text{nb-ue}} \tag{1}$$

Scenario 2: only through the channel of RIS, the direct channel is blocked.

$$H_T^{'} = H_{\text{ris-ue}} \Theta_{\text{ris}} G_{\text{nb-ris}} \tag{2}$$

Correspondingly, the received signals $Y$ at UE can be expressed as follows:

Scenario 1: both direct channels and channels passing through RIS are included

$$Y = (H_{\text{ris-ue}} \Theta_{\text{ris}} G_{\text{nb-ris}} + H_{\text{nb-ue}}) FX + W \tag{3}$$

Scenario 2: only through the channel of RIS, the direct channel is blocked.

$$Y^{'} = (H_{\text{ris-ue}} \Theta_{\text{ris}} G_{\text{nb-ris}}) FX + W \tag{4}$$

Where $\Theta_{\text{ris}} = \text{diag}(\theta_1, \theta_1, \cdots, \theta_N)$ is the reflection vector at RIS, $\theta_n$ represents the reflection coefficient of the $n_{\text{th}}$ RIS elements, $\beta$ is RIS gain, $F$ is the preencoding vector at NB, $W \in C^{U \times 1}$ is the receiving noise at UE. Note that $\theta_n$ can be further expressed as $\theta_n = \beta_n e^{j\varphi_n}$, $\beta_n \in [0,1]$ and $\phi_n \in [0, 2\pi]$ and represent the amplitude and phase of the $n_{\text{th}}$ RIS element, respectively.

3.2 Problem analysis

Scenario 1: both direct channel and scattering channel through RIS are included

It can be seen from the above that the channel through RIS includes two segmented subchannels: NB-RIS and RIS-UE. Typically, the RIS panel is generally arranged at a relatively high position. Typically, there are few scatterers between Nb and RIS. Most channels between NB-RIS are pure LOS channels or rice channels with high rice factor, and the channel matrix is in a low rank state. According to the inequality $r(AB) \leq \min(r(A), r(B))$ satisfied by the rank of the product of the two matrices, even if the channel between RIS-UEs is a high rank channel with rich scattering, the rank of the total channel of NB-RIS-UE is still limited by the channel between NB-RIS. The characteristics of RIS channel are similar to that of multi antenna relay channel. The low rank problem of multi antenna relay channel is deeply analyzed in reference [16]. In addition, it can also be analyzed from the perspective of keyhole effect of MIMO channel. After the introduction of RIS, the RIS channel may produce keyhole effect, especially in the scenario where the NB-UE direct channel is blocked and only the channel is reflected through RIS [17].

Scenario 2: the direct channel is blocked only through the RIS scattering channel

From the formula (1), the channel component of RIS includes the beamforming gain $\beta$ provided by RIS. Typically, RIS generally has a very large-scale antenna array, that is, the beamforming gain is high ($\beta \gg 1$), so the contribution of the channel component $H_{ris-ue}\beta\Theta_{ris}G_{nb-ris}$ through RIS is much greater than the channel component $H_{nb-ue}$ from the base station to the UE. In this typical case, scenario 1 degenerates to scenario 2, and the number of channel conditions is mainly limited by the channels between NB-RIS. Of course, if $\|H_{nb-ue}\| \gg \|H_{ris-ue}\Theta_{ris}G_{nb-ris}\|$ makes the magnitude of $\beta\|H_{ris-ue}\Theta_{ris}G_{nb-ris}\|$ equal to $\|H_{nb-ue}\|$, that is, the contribution of channel components after RIS is equal to that of base station directly to UE channel. At this time, the number of channel conditions is similar to that of traditional MIMO channel without RIS.

3.3 Potential solutions

Based on the analysis of the above problems, the restriction on the number of channel conditions after the introduction of RIS mainly comes from the channel $G_{nb-ris}$ between NB-RIS. It is necessary to explore solutions from the $G_{nb-ris}$ component in the RIS segmented channel.

The cell edge UE is power limited. The main purpose of introducing RIS is to obtain array gain rather than spatial multiplexing gain. The reduction of channel rank caused by RIS has little impact on its performance. Moreover, the introduction of RIS also obtains beamforming gain and enhances the signal, so the edge UE benefits from RIS. However, for non cell edge UEs, the signal strength is generally high without RIS assistance, and how many typical natural propagation channels meet the rich scattering condition, which is a high rank channel, which can obtain high MIMO multiplexing gain. However, if RIS is introduced, that is, NB-RIS channel is introduced, the rank of the channel may be sharply reduced, resulting in a significant reduction in the MIMO multiplexing gain of the UE. Based on this qualitative analysis, the UEs with high SINR in the cell center have a greater impact on the performance of RIS, and have a relatively small impact on the performance of UEs distributed in other locations. Therefore, we mainly focus on the UE with high SINR, and further discuss the corresponding solutions in the near-field and far-field.

If the channel conditions between the base station and RIS meet the near-field assumption, Los MIMO technology can be used for analysis. For example, the phase difference of the incident signal on different antenna elements on the RIS surface can be increased by shortening the distance between the base station and the RIS and expanding the aperture area of the RIS antenna, so as to improve the number of channel conditions. Further, if the channel conditions of orbital angular momentum (OAM) are satisfied (i.e. near-field and quasi coaxial), vortex angular momentum can be considered to improve the spatial multiplexing gain. However, if the OAM mechanism is adopted, the antenna configuration of the base station and UE needs to meet the OAM requirements, and the transceiver also needs to be optimized accordingly.

If the channel conditions between the base station and the RIS meet the far-field assumption and there are only channels reflected through the RIS, it is obviously difficult to improve the number of channel conditions by relying on a single RIS panel. If conditions permit, multiple RIS panels can be deployed to provide multiple independent scattering surfaces, which can improve the number of channel conditions. The low cost, low power consumption and easy deployment features of RIS provide the possibility of deploying multiple RIS panels. If more RIS panels can be deployed, there is even a chance to solve the problem of large number of channel conditions caused by sparse scatterers in natural scattering environment, increase the degree of freedom of channel space, and then improve the spatial multiplexing gain of channel.

For the case where NB-RIS-UE channel and NB-UE channel exist at the same time, for those non SINR limited UEs, RIS can adopt wide beam or broadcast beam to reduce beamforming gain, so as to meet the condition that $\beta \| H_{\text{ris-ue}} \Theta_{\text{ris}} G_{\text{nb-ris}} \|$ is equivalent to $\| H_{\text{nb-ue}} \|$, that is, the component contribution through RIS channel is equivalent to that through base station to UE channel. At this time, RIS provides a certain beamforming gain, but also provides more scattering paths, increases the spatial degree of freedom of the channel, and then improves the spatial multiplexing gain. That is, the RIS design goal of the community center is not to pursue the shape enhancement signal, but focuses on suppressing the multipath effect, increasing the spatial degree of freedom and optimizing the stability of signal quality in different geographical regions.

In addition, in order to better select appropriate solutions for different situations, some measurement and decision mechanisms need to be designed to accurately estimate and judge the scene of UE, so as to adopt the corresponding optimization scheme. For example, if there are both RIS path and direct path, give appropriate decision criteria and select the path or direct path to schedule through RIS. Alternatively, the RIS deployment is optimized so that the cell center UE uses the direct path instead of the RIS path as much as possible.

## 4 Inter networks - multi-network coexistence

In the actual networking scenario of wireless mobile communication network, multi network coexistence is a traditional typical problem. Multiple coexisting networks are deployed and managed by different operators. The key challenge of coexistence is the limited cooperation between them. After the introduction of RIS, it can optimize the interference relationship between adjacent nodes under ideal regulation. However, considering the complexity of implementation, the feasibility of multi network coordination and the inherent technical characteristics constraints of RIS, it brings new challenges to multi network coexistence. In the actual network, the wireless signals incident on the RIS panel include both "target signals" optimized and regulated by RIS and other "non-target signals". RIS will regulate these two kinds of signals at the same time. By adjusting the amplitude, phase and polarization mode of electromagnetic wave, it can enhance the "target signals" and unexpected abnormal regulation of "non-target signals".

From the perspective of spectrum usage relationship, multi network coexistence can include two scenarios: CO frequency coexistence (also known as co channel coexistence) and different frequency coexistence (also known as adjacent channel coexistence). When multiple networks coexist, the main restrictive factor is that the networks cannot or are inconvenient to coordinate regulation / scheduling. This section will analyze the challenges faced by these two types of coexistence scenarios respectively.

4.1 Co-frequency coexistence

With the increasing shortage of spectrum resources, spectrum sharing will be the main spectrum use mode of 6G networks in the future [3]. When multiple networks use spectrum by spectrum sharing, there is a problem of CO frequency coexistence among networks. Traditional spectrum sharing technology has been very mature. Typical methods include spectrum sharing between master-slave systems based on cognitive radio (CR), unauthorized spectrum sharing technology based on IEEE 802.11 standard, etc. The introduction of RIS will provide new opportunities for spectrum sharing technology, but it will also bring new challenges.

(1) Unexpected abnormal regulation of network signals with the same frequency and different frequency

The typical coexistence of multiple networks is usually a common coverage deployment. The overlapped coverage of the deployed RIS will simultaneously control the same frequency signals coming from different networks. In case 1, it is assumed that two networks with the same frequency (network a and network B) coexist, and it is assumed that only network a has RIS-A deployed. RIS-A is controlled by the network A, and the target signal coming from the network A is optimized and regulated. It also has unexpected abnormal regulation on the same frequency non target signal coming from the B network. In case 2, it is assumed that two networks with the same frequency (network a and network B) coexist, and it is assumed that networks a and B deploy RIS-A and RIS-B respectively. RIS-A is controlled by network a, and the target signal from network a incident on it is optimized and regulated. At the same time, RIS-A will also perform unexpected abnormal regulation on the same frequency non target signal from network B incident on it. Accordingly, RIS-B is controlled by network B, and the target signal from network B incident on it is optimized and regulated. At the same time, RIS-B will also perform unexpected abnormal regulation on the same frequency non target signal from network a incident on it

For case 1, the signal propagation channel of network B produces unexpected dynamic changes. If the channel $H_B(t_1)$ measured by network B at time $t_1$ is abnormally regulated by RIS-A at time $t_2$, the channel $H_B(t_2)$ at this time may be very different from $H_B(t_1)$. Network B may schedule transmission at $t_2$. Because network B does not know the change of its channel, it still schedules based on $H_B(t_1)$, and the channel mismatch leads to serious performance degradation; In case 2, because network a and network B may be regulated by their own RIS optimization and the other party's RIS abnormal regulation at the same time, the RIS regulation and optimization may fail, and even the performance may deteriorate. In addition, the introduction of super large antenna aperture RIS also makes the non-stationary characteristics of MIMO channel more significant [18], which will further aggravate the impact of RIS abnormal regulation.

There are two possible solutions to the unexpected abnormal regulation of the same frequency signals of different systems / networks. First, for the scenario of RIS dynamic regulation, reduce the period of channel measurement and reduce the probability of channel mismatch caused by RIS as much as possible; Second, when multiple networks share and use the spectrum, if the random competitive access mechanism based on channel perception such as "listen before talk (LBT)" is adopted, the TDM mode will naturally be formed between the same covered networks to use the spectrum, so that the signals of network a and network B will not enter the RIS and be abnormally regulated at the same time.

(2) RIS strengthens the signal and expands the coverage radius, but it will also deteriorate the interference relationship between networks

Spectrum sharing between networks is generally based on random competitive access of interference energy sensing to realize fair spectrum sharing between systems. The introduction of RIS brings great changes in channel characteristics, which no longer meets the wide stationary assumption, which challenges the accuracy of spectrum interference sensing and evaluation. In RIS network, the interference signals to the surrounding nodes not only come directly from the transceiver nodes, but also the interference signals enhanced by RIS beamforming. RIS performs dynamic abnormal regulation on the signal, which makes it difficult to accurately estimate the signal of the path, and the signal strength of the path is high, and its estimation deviation will have an impact that can not be ignored. In addition, a typical RIS has a large-scale antenna array element and a large-scale antenna aperture, and the spatial non-stationarity is more significant, which poses a greater challenge to the accuracy of spectrum interference sensing and evaluation. The increase of interference sensing and estimation bias will aggravate the hidden node / exposed node problem in spectrum sharing, which may worsen the coexistence relationship between systems.

The interference perception and estimation deviation caused by RIS beamforming is mainly caused by the shaped directional narrow beam. The traditional omnidirectional antenna assumption can not accurately measure the interference signal, and the LBT mechanism in spatial dimension needs to be considered. In our previous research, we provided a directional LBT mechanism, which can more accurately sense the interference signal of narrow beam [19]. Of course, other conservative schemes can also be adopted. For example, reduce the sensing threshold in RIS network to avoid interference as much as possible; Or, during RIS regulation, add optimization constraints, that is, limit its coverage after signal regulation, and optimize the signal of the target system on the premise of ensuring coexistence performance. However, the spectrum efficiency of these conservative schemes is low and the system performance will be lost.

To sum up, in the spectrum sharing scenario, the basic technical idea to solve the multi-networks coexistence after the introduction of RIS is to optimize the traditional spectrum sharing access mechanism, such as adopting the directional LBT mechanism and reducing the sensing threshold, so as to reduce the probability of hidden nodes / exposed nodes as much as possible. Through the optimized competitive access mechanism, the spectrum sharing and use relationship between TDM mode networks is achieved.

4.2 Coexistence of different frequencies

For different frequency coexistence scenarios, there is a contradiction between the broadband regulation of RIS and the requirements of multi-networks adjacent channel leakage suppression. In this section, the problem of RIS under different frequency coexistence will be deeply analyzed and the corresponding solutions will be given.

Conventional RIS generally has the ability of broadband regulation, which is conducive to wireless broadband communication and supports the communication of multiple frequency bands at the same time. However, this feature is in contradiction with the coexistence of multiple network adjacent frequencies. The antenna array element of RIS can only regulate a single phase/amplitude at the same time, and can not use different coefficient weighting for the incoming signals of different subbands at the same time, so it is impossible to carry out the best channel matching for more than one subband channel. For example, two networks (network a and network B) overlap and cover, and both use the spectrum resources of adjacent frequency bands. If the RIS-A of network a optimizes and regulates the signal propagation according to its communication requirements, since the conventional RIS is broadband electromagnetic wave regulation, RIS-A will simultaneously regulate and control the incident signals within a large bandwidth, that is, it will also regulate and control the signals of adjacent frequency network B in the same coverage area, resulting in serious unexpected channel disturbance to different frequency network B, It brings the coexistence problem between different frequency networks. That is, the broadband regulation of RIS conflicts with the requirements of multi network adjacent channel leakage suppression. For example, operators a and B are adjacent to each other, and RIS-A optimizes the regulation of electromagnetic wave propagation for operator a; Since the general RIS regulation response bandwidth is wide (for example, a few GHz bandwidth), RIS-A will also regulate the wireless signal of operator B. Obviously, the regulation of operator B signal by RIS-A is unexpected, so it may have a serious impact on the performance of operator B.

One possible solution to the unexpected regulation problem of different frequency coexistence scenario is RIS with multi-layers meta-surface structure. Without losing generality, here we take the double-layer super surface structure as an example. Among them, the first layer is the transmission surface, which generally adopts a fixed weight to limit the band of the incident signal, only pass through the signal within the target bandwidth, and suppress the out of band signal; The second layer is the normal dynamic adjustable meta-surface, which regulates the signal in the filtered target bandwidth. It should be noted that the first super surface of the reflective double-layer structure RIS will perform band limiting filtering on the signals outside the adjacent frequency band twice, that is, the adjacent frequency band signals will be filtered when the signals are incident and emitted from the inside. If more layers of band limiting meta-surfaces are used for band limiting filtering, the band limiting effect will be better. However, the effects of inter-layer coupling, fading of target signal, cost and volume caused by multi-layers structure need to be evaluated.

It should be noted that on the premise of ensuring the regulation performance of the target signal, improving the band limited performance index will bring complexity and cost problems. however,considering the relatively low quantization accuracy requirements of RIS itself,it is possible to reduce the impact index of out of band leakage in principle Further in-depth analysis and evaluation are needed to seek the optimization scheme on the basis of balancing the relationship between performance and cost.

4.3 Coexistence of private network and public network

A key design goal of 5g network is to support the application of vertical industry, and it can be expected that 6G will continue to strengthen its support for vertical industry in the future. Vertical industry users and ordinary users face different needs. Using private network to serve vertical industry users is a typical way. With the rapid development of vertical industry users, private networks will also exist widely. Therefore, the coexistence of private networks (serving vertical industries) and public networks (serving ordinary users) can not be ignored. Although the coexistence of private network and public network can also belong to the same frequency / different frequency coexistence problem mentioned above, it has some unique characteristics, which are discussed here.

There are two ways to realize private network coverage. They face different challenges and need to adopt different solutions. Private network mode 1 adopts the public network to realize the private network coverage requirements (one network), that is, a public network is used to meet the needs of private network users and other general users in the same coverage area at the same time. RIS regulation and resource scheduling give priority to private network users. When private network users have business needs, RIS regulation gives priority to optimizing private network user coverage; If there is no private network service demand, the demand of other users covered by the network shall be given priority. Private network mode 2, independent private network coverage, serves private network and public network users respectively with the public network. The typical situation is that the two are different frequency networks, and there is no co frequency coexistence problem in mode 1. However, if the private network is adjacent to the public network, it will face the problem similar to the coexistence of different frequencies in Section 4.2, and the solution is also similar. However, considering the high priority of general private network services, it is necessary to take into account the optimization requirements of private network services with high priority when adopting the solution of different frequency coexistence.

## 5 Intra-networks - multi-user multiplexing and multi cell coexistence

There are two kinds of coexistence relationships in a cellular network, including multi UE coexistence in cells and inter cell frequency reuse / adjacent channel coexistence. The introduction of RIS into the network can achieve better interference control under ideal regulation. However, considering the implementation complexity and the inherent technical characteristics of RIS, there are great challenges to the coexistence of the above two relationships in practical engineering applications.

5.1 Multi-UE multiplexing

For OFDM system, different UEs in the cell allocate resources by frequency division multiplexing (FDM), which is a typical multi UE multiplexing method, that is, different UE allocation occupies different frequency domain subbands (different frequency domain resource blocks (RBs)). As mentioned above, when RIS performs electromagnetic wave regulation, the minimum frequency domain bandwidth regulated at the same time is wide, which is generally much larger than the system bandwidth of the cellular cell. Therefore, only a certain phase / amplitude modulation can be performed at the same time within the system bandwidth of the cellular cell. Different matrix weights cannot be used for each frequency domain subband of the frequency division multiplexing UE at the same time, and the channels of different UEs cannot be best matched. Therefore, if FDM is used to obtain the degree of freedom gain in frequency domain, there may be no way to optimize the degree of freedom in spatial domain at the same time. This means that even if FDM scheduling is adopted, these UEs have to assume that RIS can only use the same precoding matrix, which limits the flexibility of multi-user scheduling.

For the multi-user FDM and RIS beam constraint problem after the introduction of RIS, one solution can learn from the processing mechanism of massive MIMO hybrid beamforming. The multi-user FDM scheduling scenario is limited by the flexibility of beamforming, which is not a new problem introduced by RIS. The analog beamforming of massive MIMO hybrid beamforming will also have similar constraints on multi-user FDM scheduling. The first mock exam encoding matrix can only be used in Massive MIMO active phased array antenna at the same time, that is, only one analog beam can be formed. Different UE can adapt the channel by using different digital beams under the same analog beam. Taking the following line link as an example, referring to massive MIMO hybrid beamforming (NB baseband digital beam + phased array antenna analog beam), the downlink channel after introducing RIS constitutes the hybrid beamforming communication model of NB + RIS. In the joint precoding optimization of downlink Nb and RIS, RIS adopts one precoding matrix at the same time, while NB side adopts different precoding matrices for different UEs to better adapt the channel. Like the constraints of the massive MIMO hybrid beamforming scenario, the above RIS scenario also requires that the simultaneously scheduled UE is a UE group that can belong to the same RIS analog beam. Taking the scenario of formula (4) as an example, the above mechanism can be expressed as formula (5). The uplink is UE + RIS hybrid beamforming, and a similar mechanism can be adopted.

$$Y_{ue\_i} = (H_{ris\text{-}ue\_i} \Theta_{ris} G_{nb\text{-}ris\_ue\_i}) F_{ue\_i} X + W ,  \quad (5)$$

Wherein, the RIS weighting matrixis the same,; is the precoding matrix on the NS side; andare the channels experienced by FF between NB-RIS and RIS-UE, respectively.

Wherein, the RIS weighting matrix $\Theta_{ris}$ is the same, $\forall i$; $F_{ue\_i}$ is the precoding matrix of $ue\_i$ on NB side; $H_{ris\text{-}ue\_i}$ and $G_{nb\text{-}ris\_i}$ are the channels experienced by $ue\_i$ between NB-RIS and RIS-UE, respectively.

Compared with active phased array antenna, generally passive RIS has larger antenna array and larger antenna aperture. Our previous research [14] provided a RIS blocking mechanism, which can divide the super large RIS surface into multiple sub blocks. Different sub blocks are used for beam regulation and shaping of incident signals of different UEs to improve the flexibility of beam shaping. Of course, blocking also reduces the effective antenna aperture size, resulting in the decrease of RIS gain. Therefore, it is necessary to weigh the gain relationship between the flexibility of beamforming and the reduction of antenna aperture.

If multiple RIS panels are deployed in the cell, different RIS can be used to regulate the incident signals of different UEs. The advantages and disadvantages of multi-RISs scheduling are similar to RIS blocking, which will not be repeated here.

Further, if multiple UEs multiplexed by FDM have different QoS requirements on delay, reliability and bandwidth, the above mechanism can be optimized accordingly. For example, for RIS blocking mechanism, UE with high QoS requirements gives priority to allocating RIS sub blocks with larger size; For sharing multiple RIS, UEs with high QoS requirements are given priority to allocate more RIS, and RIS with better distribution location is selected. When multiple RIS are allocated, the RIS allocated by different UEs may be overlapped or non overlapped. In the case of overlap, the RIS of the overlapping part is optimized and partitioned according to the business priority.

## 5.2 Multi-cell coexistence

Similar to multi-networks coexistence, multi-cell/node coexistence in the network also includes co-frequency coexistence (frequency reuse) and different frequency coexistence (different carriers are used between cells). Therefore, the above scheme for multi network coexistence can also be used for multi cell coexistence. However, compared with multi-networks coexistence, there is an opportunity for cooperative optimization between adjacent transmission nodes in the network. In other words, the nodes in the multi-networks are non cooperative coexistence relations, while the different nodes in the network are cooperative coexistence relations.

For the target signal, there is a cooperative coexistence relationship between different nodes in the network, which is similar to the CoMP and cell-free networks. The relevant cooperative interference suppression and signal cooperative transceiver mechanisms can be used for reference. However, RIS provides new flexibility: ①when different cells cooperate to share RIS, they only need to interact with RIS regulation signaling with low bandwidth requirements, rather than data with high bandwidth and high real-time requirements; ②More deployment quantity and larger size of RIS provide the flexibility of RIS resource scheduling; ③RIS can not only enhance the signal through beamforming, but also suppress the amplitude absorption signal, and can regulate the phase / polarization abnormally, which provides more possibilities for inter cell cooperation and sharing RIS.

With real-time dynamic cooperation and interaction of channel state information, RIS can isolate the interference relationship between adjacent nodes of the same frequency as much as possible. For example, isolate the signal from the adjacent RIS/AP, that is, suppress the signal amplitude from the incident angle of the adjacent node or completely absorb the signal without scattering, and only optimize the signal regulating the incident direction of the signal in the cell. Through different regulation and processing of signals with different incident angles, the same frequency interference of adjacent nodes is suppressed as much as possible, and the isolation between nodes is improved, so as to improve the frequency reuse coefficient of cellular network.

RIS can be shared among cells through close cooperation. Mode 1: inter cell cooperation, semi-static time slot division or dynamic time sharing, using the same RIS; In mode 2, the RIS surface is partitioned, and different cells coordinate to use different RIS sub blocks. The mechanism is similar to that described above. Multiple UE blocks use one RIS.

Use RIS to realize the flexible scaling of cell coverage (cell breathing). By flexibly regulating the RIS in the cell, especially the RIS deployed at the edge of the cell, the beam forming gain and beam coverage area of the cell signal are controlled, so as to realize the flexible scaling regulation of the cell coverage. However, the scaling of the coverage of a cell will have an impact on the surrounding adjacent cells, so close cooperative coverage relationship between adjacent cells is required.

In the future, 6G network may break the topology of traditional cellular network and introduce a new de cellular structure. Reference [20] discusses RIS assisted de cellular networks in typical broadband scenarios, and proposes a joint precoding framework of BS and RIS to maximize network capacity. For the non convexity and high complexity of joint precoding, the authors propose an alternating optimization algorithm to solve this challenging problem.

For the non target signal, there is a problem of unexpected abnormal regulation similar to the multi UE

multiplexing described above, and the solution idea is basically the same, except that the relationship between different UEs in the cell becomes the relationship between different UEs in the cell, which will not be repeated here. However, compared with different UEs in the cell, the real-time and flexibility of inter cell cooperation will be limited. Generally, the mechanism with relatively low requirements for real-time cooperation is adopted.

To sum up, the adjacent transmission nodes in the network have the opportunity to cooperate and use RIS. In order to realize the cooperation among multiple cells, it will also face the problem of complexity. In practical networks, it is necessary to balance the relationship between collaboration complexity and coexistence performance.

# 6 RIS network deployment

With the characteristics of low cost, low power consumption, simple and easy deployment, RIS has the opportunity to be widely deployed in the network, intelligently regulate the electromagnetic propagation environment, and bring a new network paradigm. However, as a newly introduced network element, RIS faces great challenges for its ubiquitous deployment in the network due to its unique technical characteristics and application scenarios. Different application scenarios may require different intelligent hypersurface deployment strategies, which need to be designed according to actual needs, such as improving coverage, reducing electromagnetic interference, improving positioning accuracy, etc. In addition, the networking design of RIS needs to consider not only the implementation scheme of the network architecture in the traditional cellular network, but also the research and exploration of the implementation scheme in the future new network architecture network, such as the networking implementation based on cell free architecture. In this section, we will try to identify the typical deployment scenarios of RIS, study the possible problems and challenges of the identified typical communication scenarios, and put forward the corresponding candidate deployment schemes.

## 6.1 Basic concepts

### 6.1.1 RIS deployment scenarios

From the perspective of communication environment complexity and RIS deployment and regulation complexity, deployment scenarios can be divided into two categories: small-scale controllable restricted area and large-scale complex environment. These two types of scenarios have great differences in RIS network deployment principles and requirements.

Small controllable restricted areas have the opportunity to deploy RIS with sufficient density and realize accurate intelligent regulation of electromagnetic environment, such as typical indoor hot spot coverage areas. In such areas, the wireless propagation environment is relatively independent, the number of main scatterers is limited, and it is convenient to deploy RIS panels on the corresponding surfaces; Generally, it is a hot area of business requirements, and more business requirements are relatively concentrated and stably distributed in this geographical area. For such areas, a sufficient number and large-size RIS can be deployed to replace the main scatterer surface in the original natural environment, and even more RIS can be deployed at appropriate locations as needed to increase the scattering surface (that is, RIS can be deployed at appropriate locations, more scattering paths can be artificially introduced, and flexible RIS selection and scattering regulation can be adopted, Realize the purpose of propagation path reselection and channel reconstruction). In a limited geographical area, RIS with sufficient density can jointly optimize scheduling and regulation, accurately regulate the wireless communication environment on demand, and build a wireless intelligent environment that can be almost accurately described and controlled. The topology of this scenario can not only suppress large-scale fading, but also realize dynamic tracking and control of phase / amplitude and Doppler frequency shift of small-scale multi-path channel through accurate regulation of RIS, so as to suppress multi-path fading effect. From the perspective of RIS morphological requirements, in order to achieve accurate regulation, RIS with dynamic adjustable capability is required. Therefore, this kind of RIS will also be higher in structure and control complexity, cost, power consumption and so on. However, such scenarios are generally hot spots, insensitive to cost, and the scope of geographical space is limited. RIS deployment and optimization are relatively simple.

Large scale complex environment, relatively sparse service distribution, inconvenient and unnecessary to achieve accurate control of wireless propagation environment. For such an environment, we can focus on the regulation of the large-scale characteristics of wireless propagation channels, including shadow fading, free space propagation path loss and other large-scale characteristics. For the scene with serious shadow blocking in natural propagation channel, RIS is deployed at the appropriate position to regulate the scattering angle and build a new propagation path of NB-RIS-UE, so as to overcome the problem of shadow blocking. As shown in Figure 1 (a), there is a tall building block between the base station and UE, and RIS can be deployed on the surface of the building to control the scattering of the signal, and a new propagation path is constructed. In addition, by deploying large-scale array element RIS to obtain high beamforming gain, the free space propagation path loss can be overcome to a certain extent. As shown in Figure 1 (b), the UE signal strength at the edge of the cell is limited, and RIS can be deployed at UE near the edge of the cell. RIS with large antenna aperture provides large antenna gain. The downlink can improve the received signal strength of UE at the cell edge, and the uplink can improve the beamforming gain of UE uplink transmitted signal. For such scenarios of channel large-scale characteristic regulation, because the channel characteristics change slowly or basically unchanged, the dynamic requirements of RIS regulation are relatively low. It can be considered to select RIS with low response rate or even fixed weight. It can be seen that for a large-scale complex environment, RIS will mainly regulate the existing or newly introduced main propagation paths / main scatterers to realize semi dynamic or static regulation of the large-scale characteristics of wireless channels. The required RIS form is simple, easy to deploy and low cost.

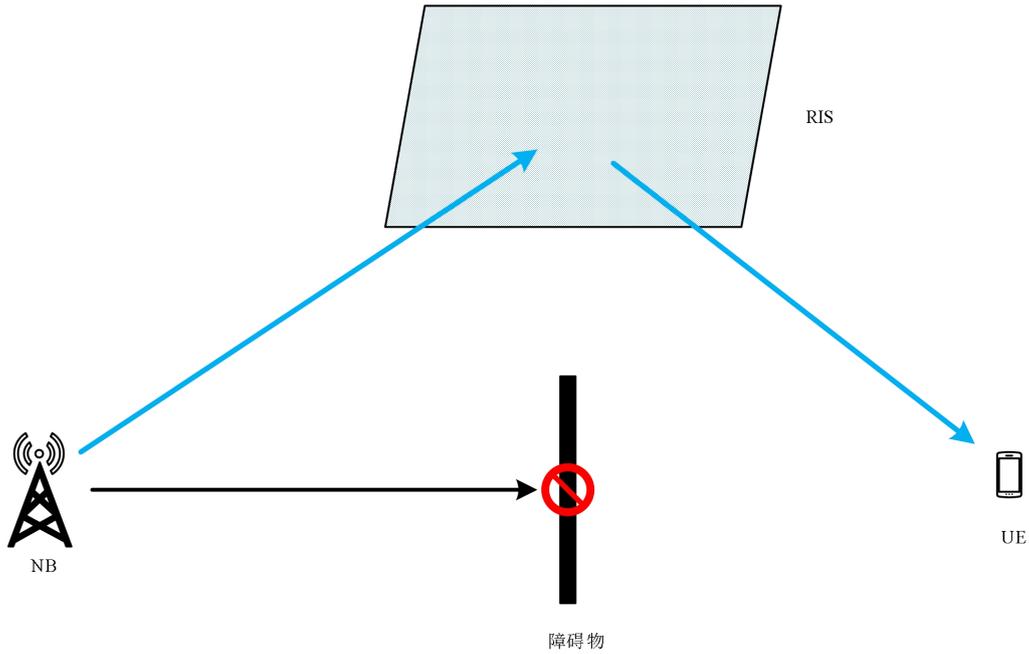

(a) NB-RIS-UE new propagation path overcomes shadow blocking

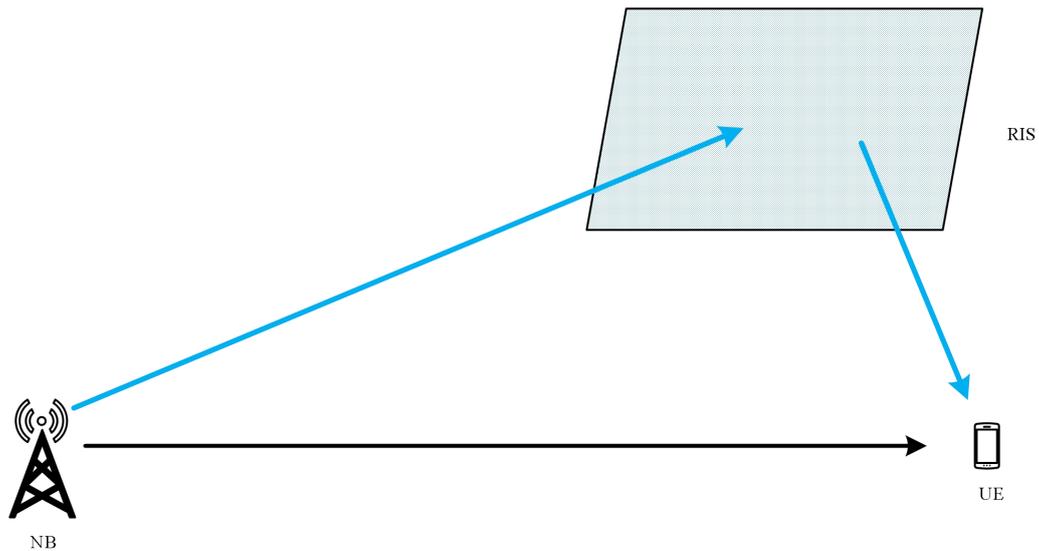

(b) Cell edge coverage enhancement

Fig.1 Typical scenarios for RIS deployment

6.1.2 Supports multi band coexistence

In order to meet the demand for higher throughput of the network in the future, the network needs to have the full spectrum capability to support the frequency band from below 6 GHz to terahertz at the same time. Traditional schemes need to deploy multiple sets of distributed medium RF units supporting multiple frequency bands to support high and low frequency bands at the same time, which is a great challenge in cost and complexity. RIS has the potential to support cross band. It has the opportunity to deploy as few RIS as possible to support the whole band, and even use only a single set of RIS to support the regulation requirements of the whole band. That is, different base stations supporting high and low frequency bands need to be deployed, while RIS only needs to deploy one set, which can greatly reduce the cost and deployment complexity. It should be noted that the cost of supporting full band RIS is high, especially supporting high-frequency RIS. However, in general, hot areas only need high-frequency coverage to support large bandwidth services, so RIS supporting high-frequency can be deployed only in these areas. In addition, different regions and operators can use different frequency bands. RIS supporting different frequency band combinations can be deployed according to the respective frequency band coverage requirements of different regions and operators, so as to balance the relationship between RIS cost and deployment complexity.

6.1.3 basic principles and processes of RIS deployment and Optimization in typical communication scenarios

The network deployment of traditional classic communication scenarios can include indoor coverage, outdoor coverage, outdoor coverage, indoor coverage, etc. RIS can be used to support blinding, weakening and increase channel freedom in these scenarios. The basic principles of network deployment can include: ① ensure that the signal strength in the coverage area is higher than the expected threshold, so as to meet the minimum transmission rate; ② Ensure that the signal strength or SINR distribution in the target coverage area is stable and avoid unexpected mutations. For the latter, we can maintain high signal strength in the coverage area through reasonable RIS deployment and regulation. Alternatively, the RIS dynamic beam tracking is scheduled to ensure the signal strength of the UE with service requirements, but the access requirements of the idle state UE anytime and anywhere need to be considered, that is, the basic coverage signal strength needs to meet the initial open-loop access. After the UE is accessed, it enters the closed-loop control. RIS can dynamically regulate the beam to track the UE, cover the connected state UE with a stronger signal, and achieve a higher service transmission rate. It should be noted that for the need to improve the coverage capacity, especially in outdoor scenes, due to the limited location where RIS can be deployed, RIS is likely to be far away from the base station, and the signal strength of the location where RIS is located is weak. Therefore, even with RIS antenna gain, the coverage can be extended for a short distance. At this time, the RIS gain has to be increased at the cost of greater complexity and cost (for example, deploying a larger antenna aperture RIS), so as to improve the ability to expand the coverage distance as much as possible.

Figure 2 shows the basic process of RIS deployment and Optimization in a typical communication scenario. Firstly, under the constraints of complexity and cost, taking the natural channel and service demand distribution in typical scenarios as the basic input, the initial RIS deployment topology is designed. Then, based on the adaptive wireless transmission regulation performance of RIS, the deployment topology of RIS is further iteratively optimized, so as to build an intelligent and controllable wireless environment. That is, the objective of RIS deployment optimization design is to seek the balance of complexity, cost and performance, and output the topology of RIS, including deployment location, density, RIS form, regulation / cooperation relationship and other parameters.

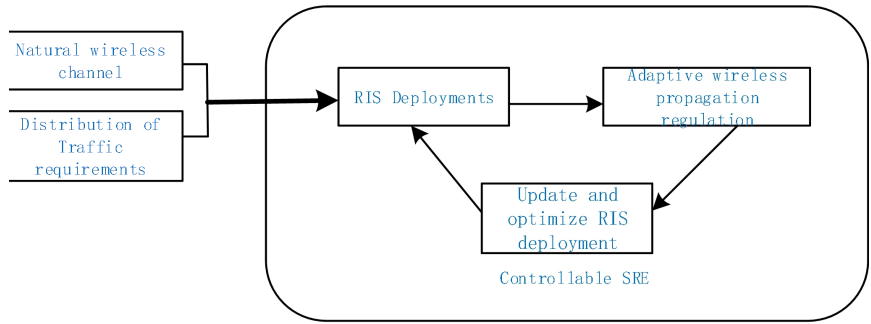

Fig.2 Basic process of RIS deployment and optimization for typical communication scenarios

6.2 An adjustment mechanism for the adaptive network capacity and coverage

In the actual scenario, the network capacity/coverage demand is unevenly distributed in geographical space, and the distribution changes dynamically or semi dynamically. For example, in large-scale events, morning and evening peaks, high-speed rail communications and other scenarios, the demand is reflected in the regular semi-static changes in different geographical spaces (the capacity demand is semi-static migration in different geographical spaces). In such scenarios, the network topology needs to be able to adaptively realize the distribution and migration of network capacity in geospatial latitude. Therefore, the network topology and regulation design need to be able to semi dynamically adjust the coverage and capacity of network resources, that is, maximize energy efficiency; Minimizing the cost is to overcome the coverage hole problem and meet the capacity requirements by minimizing the number of base stations and cells. According to the above demand characteristics, under cost constraints, we provide an example of adaptive network capacity and coverage adjustment based on RIS.

The first step is to make semi-static adaptive adjustment to the network coverage according to the imbalance and semi-static change of coverage and capacity demand in geographical regions. The specific implementation can include three types of options: ① use the movable UAV and high-altitude platform to carry Nb, and adjust the capacity / coverage of the air platform in a large physical range; ② Adopt unmanned / manned vehicles to carry ground platforms such as Nb / relay, and adjust the capacity / coverage of large geographical areas; ③ Higher altitude platforms can use satellites to carry Nb and make coverage adjustment in a larger geographical range. Among them, the first two methods can also use RIS to replace Nb and put it on UAV, high-altitude platform or unmanned vehicle to control coverage through RIS regulation signal. In the specific engineering implementation, some fixed candidate locations can be optimized, that is, the geographical space can be limited, so as to reduce the complexity of engineering implementation.

The second step is to use CoMP/cell-free mechanism to realize small-scale coverage and capacity adjustment. CoMP/cell-free can adaptively adjust the cooperative AP set and the coverage area of logical cells. Generally, the AP set of comp collaboration is small, while cell free can achieve a wider range of AP set collaboration. Introducing the RIS network and adopting the idea similar to CoMP/cell-free, it can regulate multiple RIS cooperative sets in a single cell and the cooperative sharing of RIS among multiple cells, so as to realize the small-scale regulation of coverage and capacity.

The third step is the adaptive adjustment of cell coverage. The cell breathes through adaptive power adjustment or RIS to adjust the coverage. The cell coverage and the presence or absence of interference to the surrounding area are controlled through the cell adaptive switch (cell on/off) or the scattering/absorption of RIS.

In the fourth step, RIS is used for large-scale and small-scale fine regulation of local wireless channels. For example, RIS semi-static control of large-scale characteristics is used to overcome coverage holes and blind filling. RIS is used to dynamically control the phase and amplitude of multi-path to suppress the multi-path effect.

Through the above adaptive network capacity and coverage adjustment mechanism based on RIS, it is possible to realize the so-called adaptive / intelligent flexible wireless network topology and build a new paradigm of wireless network topology.

In order to realize the above adaptive network coverage and capacity adjustment mechanism based on RIS, the following aspects need to be specially studied:

① The impact of coverage / capacity migration on spectrum allocation and sharing.

② Cell migration requires adaptive adjustment of network topology.

③ Cell ID adaptive planning.

④ Adaptive modulation of other network resources. For example, computing resources.

⑤ The demand of NB/RIS mobility on the backhaul link. For example, a wireless backhaul link may be used.

⑥ NB/RIS mobility demand for power supply. For example, the charging support capability is provided by deploying the charging station site platform of UAV base station with appropriate location and density. Alternatively, provide enough temporary docking platforms (mobile platform post stations) as fixed coverage demand stations. These platforms have space for docking UAV base stations, which can provide power supply / charging capacity, return capability, etc.

## 7 Conclusion

As one of the most potential 5G-adv and 6G key technologies, RIS has developed rapidly in academic research in recent years. By building an intelligent and controllable wireless environment, RIS will bring a new communication network paradigm to 6G in the future to meet the needs of future mobile communication. The simplified version of RIS will have the opportunity of initial commercial deployment and standardization in 5g / 5G adv stage, especially to improve 5g millimeter wave coverage. RIS enabled 6G network still faces many technical problems, deployment problems and challenges of standardization process in the future, which requires in-depth research and comprehensive evaluation of RIS key technologies and schemes, especially key technical challenges such as channel rank reduction, inter network coexistence, intra-network coexistence and network deployment. Limited by space and research depth, this paper only makes a preliminary analysis on the challenges faced by RIS engineering technology research and engineering application, and the solutions given are only qualitatively analyzed and discussed. Related issues also need to continue to conduct in-depth theoretical analysis and simulation evaluation to further verify the feasibility and performance ceiling of the scheme, so as to lay the foundation for the final industrial implementation of RIS.

Biographies

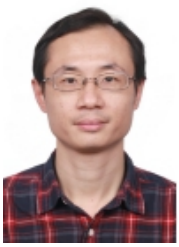

Zhao Yajun, chief engineer, Wireless Research Institute of ZTE Corp. At present, he is mainly engaged in the research of 5g standardization technology and future mobile communication technology (6g). Main research interests: reconfigurable intelligent surface, spectrum sharing, terahertz communications and flexible duplex. Email：zhao.yajun1@zte.com.cn

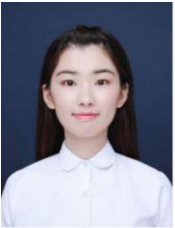

Jian Mengnan, technical pre-research engineer, Wireless Research Institute of ZTE Corp. Main research interests: reconfigurable intelligent surface, terahertz communications, holographic MIMO and orbital angular momentum. Email：jian.mengnan@zte.com.cn